\documentclass[fleqn,usenatbib]{mnras}
\usepackage{graphicx}
\usepackage{amsmath}
\usepackage{amssymb}
\usepackage{multicol}
\usepackage{bm}
\usepackage{pdflscape}
\usepackage[T1]{fontenc}
\usepackage{ae,aecompl}
\usepackage{xcolor}

 \pubyear{2020}

\begin{document}

\title[Texture in the neutron superfluid]{Vortex lattice in rotating neutron
spin-triplet superfluid}
\author[Lev B. Leinson]{Lev B. Leinson$^{1}$\thanks{%
E-mail: leinson@yandex.ru} \\
%EndAName
$^{1}$Pushkov Institute of Terrestrial Magnetism, Ionosphere and Radiowave
Propagation of the Russian Academy of Science (IZMIRAN), \\
108840 Troitsk, Moscow, Russia}
\date{Accepted XXX. Received YYY; in original form ZZZ}
\maketitle

\begin{abstract}
In the Ginzburg-Landau limit, a possible texture without singular cores in a
rotating neutron spin-triplet superfluid is studied. It is constructed the
lattice of non-singular vortices with a vorticity diffusely distributed over
the entire unit cell. The upper limit of the free energy associated with
this structure is estimated, and the result shows that non-singular vortices
are more preferable in the core of rotating neutron stars than ordinary
linear singular vortices. The order parameter of the studied neutron
system belongs to the unitary class. This implies that the superfluid under
consideration does not have spin polarization and the core-less 
$^{3}$P$_{2}$ vortices do not have magnetization, which makes electron scattering
by vortices ineffective. For the same reason, pinning with flux tubes is
also suppressed. This can affect the hydrodynamics of superfluid neutron
stars.
\end{abstract}

\label{firstpage} \pagerange{\pageref{firstpage}--\pageref{lastpage}}

%%%%%%%%%%%%%%%%%%%%%%%%%%%%%%%%%%%%%%%%%%

%%%%%%%%%%%%%%%%%%%%%%%%%%%%%%%%%%%%%%%%%%

%%%%%% BODY OF PAPER %%%%%%%%%%%%%%%%%%

\section{Introduction}

%%%%%%%%%%%%%%%%%%%%%%%%%%%%%%%%%%%%%%%%%%%%%%%%%%%%%%%%%%%%%%%%%%%%

When considering superfluid neutrons in rotating neutron stars, it is
usually assumed that the superfluid neutron flux is everywhere irrotational.
This approximation means that the velocity of superfluid neutrons $\mathbf{v}%
_{s}=\hbar \boldsymbol{\triangledown }\Phi /2M$ ($M$ is the neutron mass) is
determined exclusively by the phase gradient of the order parameter $\Psi
=\left\vert \Psi \right\vert e^{i\Phi }$ and thus $\boldsymbol{\triangledown 
}\times \mathbf{v}_{s}=0$ in complete analogy with Bose-Einstein
superfluids. In the equilibrium state, a superfluid liquid in a large
rotating container makes a rigid body rotation as a whole, which is achieved
by the presence of a lattice of quantized vortex lines. For a superfluid $%
^{4}$He this was for the first time proved by \citet{Tkachenko:1965} who has
shown that a triangular lattice of singular vortices is energetically the
most favourable of all simple lattices.

The relations, derived for a superfluid $^{4}$He in a rotating vessel are
believed valid also for a neutron Fermi liquid in neutron stars %
\citep{Richardson:1972,Muzikar:1980,Sauls:1982,Mendel:1991}). Obviously,
this statement is true in the case of spin-singlet neutron pairing with the
total spin of the Cooper pair $S=0$. In this case the condensate wave
function has the form $\Psi _{S=0}=i\sigma _{2}\left\vert \Delta \right\vert
e^{i\Phi }$, where $\boldsymbol{\sigma }=\left\{ \sigma _{1},\sigma
_{2},\sigma _{3}\right\} $ are the Pauli spin matrices while the space part
is a complex scalar function (for a recent review, for example, see %
\citep{Haskell:2017lkl,Sedrakian:2018ydt}). One again has $\mathbf{v}%
_{s}\propto \nabla \Phi $ and $\mathbf{\nabla \times }\mathbf{v}_{s}=0$,
similar to superfluid $^{4}$He in the rotating bucket.

The spin-singlet neutron pairing is believed to take place in the outer part
of the neutron star core and in the inner crust \citep{ch08} while, in the
inner core, the Cooper pairing of neutrons occurs\footnote{%
We neglect a small contribution of tensor forces.} with the orbital momentum 
$L=1$ and the total spin $S=1$. This significantly complicates the behavior
of superfluid neutrons. By this reason, in many pulsar glitch models, the
S-wave and P-wave superfluids are treated as different entities %
\citep{Ho:2015vza,montoli20}. Due to the strong spin-orbit interaction, a $%
^{3}$P$_{2}$ state with the total momentum $J=2$ is realized %
\citep{Tamagaki:1970,Takatsuka:1972}. The wave function of a pair with
momenta $\mathbf{p}$ and $-\mathbf{p}$ can be written \emph{in the form} %
\citep{Mermin:1973,Fujita:1978,Muzikar:1980,Volovik:1984} (summation over
repeated indices implied): 
\begin{equation}
\Psi =i\left\vert \Delta \right\vert \sigma _{i}\sigma _{2}\mathsf{A}_{ij}%
\hat{p}_{j},  \label{Psi}
\end{equation}%
where $\mathbf{\hat{p}}=\mathbf{p}/p$. The order parameter is a complex
matrix $\mathsf{A}_{ij}$, which, due to the total momentum constraint $J=2$,
must be symmetric and traceless: $\mathsf{A}_{ij}=\mathsf{A}_{ji}$ and $%
\mathsf{A}_{ii}=0$. This matrix is normally described in a local coordinate
system in which the real part is diagonal and the off-diagonal part is
purely imaginary \citep{Richardson:1972,Sauls:1982,Masuda:2015}.

It is believed \citep{Richardson:1972,Sauls:1982,Masuda:2015} that the
minimal energy corresponds to the order parameter of the form 
\begin{equation*}
\mathsf{A}_{ij}=Ne^{i\Phi }\left[ \hat{u}_{i}\hat{u}_{j}+\lambda \hat{v}_{i}%
\hat{v}_{j}-\left( 1+\lambda \right) \hat{w}_{i}\hat{w}_{j}\right] ,
\end{equation*}%
associated with the triad of unit mutually orthogonal vectors $\left( \hat{u}%
,\hat{v},\hat{w}\right) $. Kinetic energy can be reduced by alignment $\hat{w%
}$ (eigenvector with the smallest eigenvalue) with $\mathbf{v}_{s}$ and by
allowing the degeneracy parameter $\lambda $ to vary in the range $%
-1<\lambda <-1/2$. The planar distribution of $\hat{w}$ in this
representation immediately leads to $\mathbf{v}_{s}\propto \nabla \Phi $ and 
$\mathbf{\nabla \times }\mathbf{v}_{s}=0$, i.e. the minimal energy
corresponds to a singular vertex with the superfluid velocity distribution
similar to that shown in Fig. \ref{fig:fig1}.

This result can be easily understood. Normally the distance $d$ between
vortices in rapidly rotating neutron stars is much larger than the coherence
length of superfluid neutrons $\xi $, and therefore the authors consider a
single vortex \citep{Richardson:1972,Sauls:1982,Mendel:1991,Masuda:2015}. It
is well known that an isolated vortex in a rotating vessel of large radius
can be stabilized only if its vorticity is concentrated in the central
non-superfluid filament of size $\xi $ (a singular vortex). An isolated
core-less vortex with a diffusely distributed vorticity are unstable. 

However, core-less vortices are well known in the $^{3}$He-A theory %
\citep{Mermin:1976,Chechetkin:1976,Volovik:1977,Fujita:1978,Volovik:1984}.
They have a vorticity diffusely distributed over the entire unit cell of the
vortex lattice. For this reason, continuous vortices cannot be isolated and
should be considered as a single periodic system.

The paper is organized as follows. In Sec. \ref{sec:Fe}, we propose a
reasonable trial state for a rotating $^{3}$P$_{2}$\ superfluid system with
coordinate-dependent directions of the triad of basis vectors written in
terms of Euler angles. This type of order parameter allows one to consider
both singular and non-singular vortices with distributed vorticity. The free
energy of this vortex state in the Ginzburg-Landau regime is derived in the
London limit. In Sec. \ref{sec:ol}, we discuss the mathematical functions
used to describe the lattice of ordinary linear singular vortices in
rotating superfluid $^{4}$He. Our analysis shows that the same approach can
be used for a lattice of singular vortices in a rotating spin-triplet
neutron superfluid, when the superfluid motion is determined by the phase
gradient of the order parameter. In Sec. \ref{sec:col}\ we consider a more
general form of the order parameter making use of the additional freedom
provided by the coordinate dependence of orientation of the basis vectors
triad in rotating spin-triplet superfluid. We construct the functions
describing a lattice of non-singular vortices with a vorticity diffusely
distributed over the entire elementary cell of the vortex lattice. Since it
is difficult to solve the minimization problem, we construct some trial
function which has the lattice symmetry. In Sec. \ref{sec:est}\ we use this
trial function to calculate the free energy per area carrying a unit
vorticity and compare it with the known energy of a lattice of linear
singular vortices. Section  \ref{sec:dis}\ contains the summary of the
obtained results. Possible consequences for the neutron star dynamics are
discussed.

\section{Free energy of the rotating $^{3}$P$_{2}$ superfluid}

\label{sec:Fe}

The diffusely distributed vorticity in a periodic system should be summed
over the entire lattice into solid-state rotation, which is possible only if 
$\mathbf{\nabla \times }\mathbf{v}_{s}\neq 0$. For the $^{3}$P$_{2}$
superfluid this can be the case due to additional degrees of freedom:
coordinate-dependent directions of the triad of basis vectors. Indeed, if
the condensate wave function is of the form of Eq. (\ref{Psi}) with the
matrix $\mathsf{A}_{ij}$ depending on the centre of mass coordinate $\mathbf{%
r}=\left( \mathbf{r}_{1}+\mathbf{r}_{2}\right) /2$ of a Cooper pair:%
\begin{equation}
\mathsf{A}_{ij}\left( \mathbf{r}\right) =\frac{1}{2}e^{i\Phi \left( \mathbf{r%
}\right) }\left( u_{1i}+iu_{2i}\right) \left( u_{1j}+iu_{2j}\right) ,
\label{Aij}
\end{equation}%
where $\mathbf{u}_{1}\left( \mathbf{r}\right) $ and $\mathbf{u}_{2}\left( 
\mathbf{r}\right) $ are real unit orthogonal vectors.

The normalized symmetric matrix $\mathsf{A}_{ij}$, as given by Eq. (\ref{Aij}%
) is unitary, $\mathsf{A}_{ij}^{\ast }\mathsf{A}_{ij}=1$. Together with Eq. (%
\ref{Psi}) it can be readily recognized as the wave function of the Cooper
pairs in the $^{3}$P$_{2}$ state with $M_{J}=2$. The vector product $\mathbf{%
l}=\mathbf{u}_{1}\times \mathbf{u}_{2}$ indicates the direction of the
quantization axis of the total angular momentum of the pair.

The current density at zero temperature is 
\begin{equation}
\mathbf{j=}\frac{\hbar }{4iM}\sum_{\mathrm{spin}}\left( \Psi ^{\dag }%
\boldsymbol{\nabla }\Psi -\Psi \boldsymbol{\nabla }\Psi ^{\dag }\right)
=\rho _{s}\mathbf{v}_{s}.  \label{j}
\end{equation}%
From Eqs. (\ref{Psi}) and (\ref{Aij}) we find the invariant defined
superfluid velocity \citep{Volovik:1984}: 
\begin{equation}
\mathbf{v}_{s}=\frac{\hbar }{2M}\left( \boldsymbol{\nabla }\Phi +2u_{1i}%
\boldsymbol{\nabla }u_{2i}\right) .  \label{vsinv}
\end{equation}%
Summation over repeated indices $i,j=1,2,3$ is understood everywhere.

The triad $\left( \mathbf{u}_{1}\mathbf{,u}_{2}\mathbf{,l}\right) $ at any
point can be specified by the Euler angles $\left( \Phi ,\beta ,S\right) $
which bring the standard Cartesian frame $\left( \mathbf{x,y,z}\right) $ to
the triad. In terms of these angles we have the matrix (\ref{Aij}) with%
\begin{gather}
\mathbf{u}_{1}=\left( \cos S\cos \beta ,-\sin S\cos \beta ,\sin \beta
\right) ,  \label{m} \\
\mathbf{u}_{2}=\left( \sin S,\cos S,0\right)   \label{n}
\end{gather}%
and%
\begin{equation}
\mathbf{l}=\left( -\cos S\sin \beta ,\sin S\sin \beta ,\cos \beta \right) .
\label{l}
\end{equation}%
Note that the order parameter (\ref{Aij})\ does not change upon simultaneous
the gradient transformation and rotation of the spin-orbital space around
the axis $\mathbf{l}$%
\begin{eqnarray}
\Phi  &\rightarrow &\Phi +\alpha   \notag \\
\mathbf{u}_{1}+i\mathbf{u}_{2} &\rightarrow &\mathsf{R}\left( \frac{\alpha }{%
2}\mathbf{l}\right) \left( \mathbf{u}_{1}+i\mathbf{u}_{2}\right) .
\label{tran}
\end{eqnarray}%
Therefore, instead of three variables (three angles specifying the
orientation of the triplet of unit vectors $\mathbf{u}_{1}\mathbf{,u}_{2}%
\mathbf{,l}$), we introduced four, artificially separating the phase
variable $\Phi $. In this case, we will assume that $\Phi $\ changes only
under the gradient transformation, and the unit vectors $\mathbf{u}_{1}%
\mathbf{,u}_{2}\mathbf{,l}$\ only change under rotation. Thus $\mathbf{u}_{1}
$\ and $\mathbf{u}_{2}$\ included in (\ref{vsinv})\ are invariant under the
gradient transformation.

The vorticity of superflow, $\boldsymbol{\triangledown }\times \mathbf{v}_{s}
$ now depends on the $\mathbf{l}$ texture through the famous Mermin-Ho
relation \citep{Mermin:1976}:%
\begin{equation}
\boldsymbol{\nabla }\times \mathbf{v}_{s}\mathbf{=}\frac{\hbar }{2M}%
e_{imn}l_{i}\left( \boldsymbol{\nabla }l_{m}\times \boldsymbol{\nabla }%
l_{n}\right) .  \label{mh}
\end{equation}%
Therefore, neutron superfluid with the order parameter (\ref{Aij}) in the
central core of a neutron stars is irrotational only when the right side of
Eq. (\ref{mh}) equals zero; for example, in the case of planar field $%
\mathbf{l~}$\emph{\ }(when the vector field $\mathbf{l}$\ is concentrated in
the plane of rotation).

To determine the equilibrium state, we have to find $\Phi $, $S$ and $\beta $%
, as well as lattice parameters that minimize free energy. Throughout this
work we take the London limit, assuming spatial variations to be slow, in
other words, we assume that non-uniform states are described by the phase $%
\Phi (r)$\ and the field of the triad, while the condensation energy and
hence $\left\vert \Delta \right\vert $\ are kept constant. Since we take
London limit, we have to consider only derivative terms in the free energy.
The general form $F$ of this contribution to the free energy density,
consistent with the spin-orbit rotation and uniform gauge invariance, near
the critical temperature in the Ginzburg-Landau approximation can be written
as%
\begin{equation}
F=\frac{\hbar ^{2}\rho _{s}}{4M^{2}}\left( \partial _{i}\mathsf{\bar{A}}%
_{kj}\partial _{i}\mathsf{A}_{kj}+\partial _{i}\mathsf{\bar{A}}_{kj}\partial
_{j}\mathsf{A}_{ki}+\partial _{i}\bar{A}_{ki}\partial _{j}\mathsf{A}%
_{kj}\right) ,  \label{dF}
\end{equation}%
where the bar means a complex conjugate; $\rho _{s}$ denotes the superfluid
density. In our approach this value is an external parameter, which can be
determined, as a function of the density and temperature, from the standard
Ginzburg-Landau theory.

Since we consider rotation of the superfluid it is convenient to use the
frame of reference rotating with the container. Then the problem of
minimizing the difference $F-\mathbf{L\Omega }~$where $\mathbf{L}$ is the
total angular momentum of the neutron liquid per unit length and $\mathbf{%
\Omega }$ is the angular velocity of the container rotation can be
effectively reduced to a problem of the superfluid motion in the external
vector potential $\mathbf{A}=M\left( \mathbf{\Omega \times r}\right) $.
Therefore, accounting for the gauge invariance we obtain the relevant free
energy by replacing $\partial _{i}$ by $\partial _{i}-2iM\left( \mathbf{%
\Omega \times r}\right) _{i}$ in (\ref{dF}).

The lattice period $d=2a$ is uniquely determined by the rotation angular
velocity. The vortex density $\eta _{v}=\omega /\pi $ in units of $1/a^{2}$
and the angular velocity 
\begin{equation}
\boldsymbol{\omega =}\frac{2Ma^{2}}{\hbar }\boldsymbol{\Omega ~}.  \label{o}
\end{equation}%
are connected by Feynman's relation \citep{Feynman:1965} 
\begin{equation}
\varkappa \eta _{v}=2\omega ,  \label{Nnu}
\end{equation}%
where $\varkappa $ is the vortex strength in units of $\pi \hbar /M$. We
restrict ourselves to considering vortices with one quantum of circulation, $%
\varkappa =1$. In this case the phase $\Phi $ increases by $2\pi $ when
going around each point of the lattice.

It is convenient to use the dimensionless variable $\mathbf{\varrho }=%
\mathbf{r/}a=\{x,y,z\}$meaning hereinafter $\boldsymbol{\nabla }=\left\{
\partial _{x},\partial _{y},\partial _{z}\right\} $. Then from (\ref{dF}) we
get%
\begin{equation}
\tilde{F}=\frac{\hbar ^{2}\rho _{s}}{\left( 2M\right) ^{2}a^{2}}\left( \bar{D%
}_{i}\mathsf{\bar{A}}_{ki}D_{j}\mathsf{A}_{kj}+\bar{D}_{i}\mathsf{\bar{A}}%
_{kj}D_{i}\mathsf{A}_{kj}+\bar{D}_{i}\mathsf{\bar{A}}_{kj}D_{j}\mathsf{A}%
_{ki}\right) ,  \label{Ft}
\end{equation}%
where $\tilde{F}$ is the free energy density in the rotating frame, and 
\begin{equation}
\mathbf{D}=\boldsymbol{\nabla }-i\,\left( \boldsymbol{\omega }\times 
\boldsymbol{\varrho }\right) \mathbf{,~\ \bar{D}}=\boldsymbol{\nabla }%
+i\,\left( \boldsymbol{\omega }\times \mathbf{\varrho }\right) \mathbf{.}
\label{D}
\end{equation}

In what follows we always use the rotating frame of reference unless stated
otherwise. In the rotating frame of reference the superfluid moves relative
a normal (non-superfluid) component with the velocity $\mathbf{v}_{s}-%
\mathbf{v}_{n}$, where $\mathbf{v}_{n}$ is the velocity of the normal matter
co-rotating with the container. From Eq. (\ref{vsinv}) we obtain the
dimensionless velocity in the rotating frame \ 
\begin{equation}
\mathbf{V\equiv }\frac{2Ma}{\hbar }\mathbf{\tilde{v}}_{s}=2\cos \beta \,%
\boldsymbol{\nabla }S+\boldsymbol{\nabla }\Phi -\boldsymbol{\omega }\times 
\mathbf{\varrho }.  \label{V}
\end{equation}%
Making use of Eqs. (\ref{m})-(\ref{V}) from Eq. (\ref{Aij}) and (\ref{Ft})
we get the derivative part of the free energy density%
\begin{gather}
\tilde{F}=\frac{\hbar ^{2}\rho _{s}}{\left( 2M\right) ^{2}a^{2}}\left\{
\left( \mathbf{Vu}_{1}-\sin \beta \mathbf{\ l}\boldsymbol{\nabla }S\right)
^{2}+\left( \mathbf{Vu}_{2}-\mathbf{l}\boldsymbol{\nabla }\beta \right)
^{2}\right.   \notag \\
+\frac{1}{2}\left( \mathbf{V}-\mathbf{u}_{1}\mathbf{\times }\boldsymbol{%
\nabla }\beta \right) ^{2}+\frac{1}{2}\left( \mathbf{V}+\sin \beta \mathbf{\ 
\mathbf{u}_{2}\times }\boldsymbol{\nabla }S\right) ^{2}  \notag \\
+\frac{3}{2}\sin ^{2}\beta \mathbf{\ }\left( \mathbf{u}_{2}\boldsymbol{%
\nabla }S\right) ^{2}+\frac{1}{2}\left( 2+\sec ^{2}\beta \right) \left( 
\mathbf{u}_{1}\boldsymbol{\nabla }\beta \right) ^{2}  \notag \\
\left. +\left( \mathbf{u}_{2}\boldsymbol{\nabla }\beta \right) ^{2}+\frac{1}{%
4}\left( 3+\cos \left( 2\beta \right) \right) \tan ^{2}\beta \,\left( 
\mathbf{u}_{1}\boldsymbol{\nabla }S\right) ^{2}\right\} .  \label{F}
\end{gather}

\section{The lattice of ordinary vortex lines}

\label{sec:ol}

Since the superfluid motion is actually two-dimensional it is convenient to
define the complex coordinate $Z=x+iy$ in the plane at right angles to the
vortices. We represent the velocity by a complex quantity $\upsilon (Z)$
whose magnitude and direction in the complex plane give the magnitude and
direction of the fluid velocity at $Z$.

Let us consider first a lattice of ordinary vortex lines in a uniform
texture, which corresponds to $\beta =const$ and $S=const$. Without loss of
generality, we can take $S=0$.\emph{\ }If we assume that there exists a
vortex at $Z=0$ then the coordinates of the other vortices will be $%
Z_{mn}=2m+2ni$, where $m$ and $n$ are arbitrary integers, $1$ and $i$ are
the half-periods of the lattice (in units of $a$). Due to the irrotational
nature of the flow in singular vortices, the complex velocity $\upsilon (Z)$
must be an analytic function \citep{Milne:1964}, which has no other
singularity in the finite plane, except for simple poles at the points $%
Z_{mn}$, where there are vortices, with identical residues. All simple
lattices of this kind can undergo rigid rotation \citep{Kiknadze:1965}. It
is known that any functions of this kind whose poles form a regular lattice
can be written in the form $\zeta _{0}(Z)=\zeta (Z)+f(Z)$, where $f(Z)$ is
any entire function and $\zeta \left( Z;g_{2},g_{3}\right) $ is the zeta
function of Weierstrass with invariants (the so-called modular forms) $%
g_{2}(1,i)$ and $g_{3}(1,i)$. For brevity, we mostly omit the arguments $%
g_{2},g_{3}$ in the Weierstrass functions below.

Since in the general case it is difficult to solve the minimization problem,
we consider some test structure in the form of a square lattice of ordinary
singular vortices. In this case \citep{Tkachenko:1965} one can put $f(Z)=0$
and the dimensionless superfluid velocity in the laboratory frame $\upsilon
\left( Z\right) \equiv \left( 2Ma/\hbar \right) \upsilon _{s}\left( Z\right) 
$ can be written as $\upsilon \left( Z\right) =i\zeta \left( \bar{Z}%
;g_{2},g_{3}\right) $. In the rotating frame of reference, instead of Eq. (%
\ref{V}), we get%
\begin{equation}
\upsilon _{r}\left( Z\right) =i\zeta \left( \bar{Z};g_{2},g_{3}\right)
-i\omega Z,  \label{vc}
\end{equation}%
where the second term represents the complex velocity $\upsilon _{n}$ of a
non-superfluid component (a rigid rotation) and the bar means a complex
conjugate $\bar{Z}=x-iy$.

To evaluate the phase function $\Phi $ (let us designate it $\Phi _{0}$ for
the case of ordinary singular vortices), we consider the expression for
complex potential \citep{Ryzhik} 
\begin{equation}
\Psi \left( Z\right) =\int \zeta (Z)dZ=\ln \sigma \left(
Z;g_{2},g_{3}\right) ,  \label{PSI}
\end{equation}%
where $\sigma \left( Z;g_{2},g_{3}\right) $ is the Weierstrass $\sigma $%
-function.

\begin{figure}
\includegraphics[width=\columnwidth]{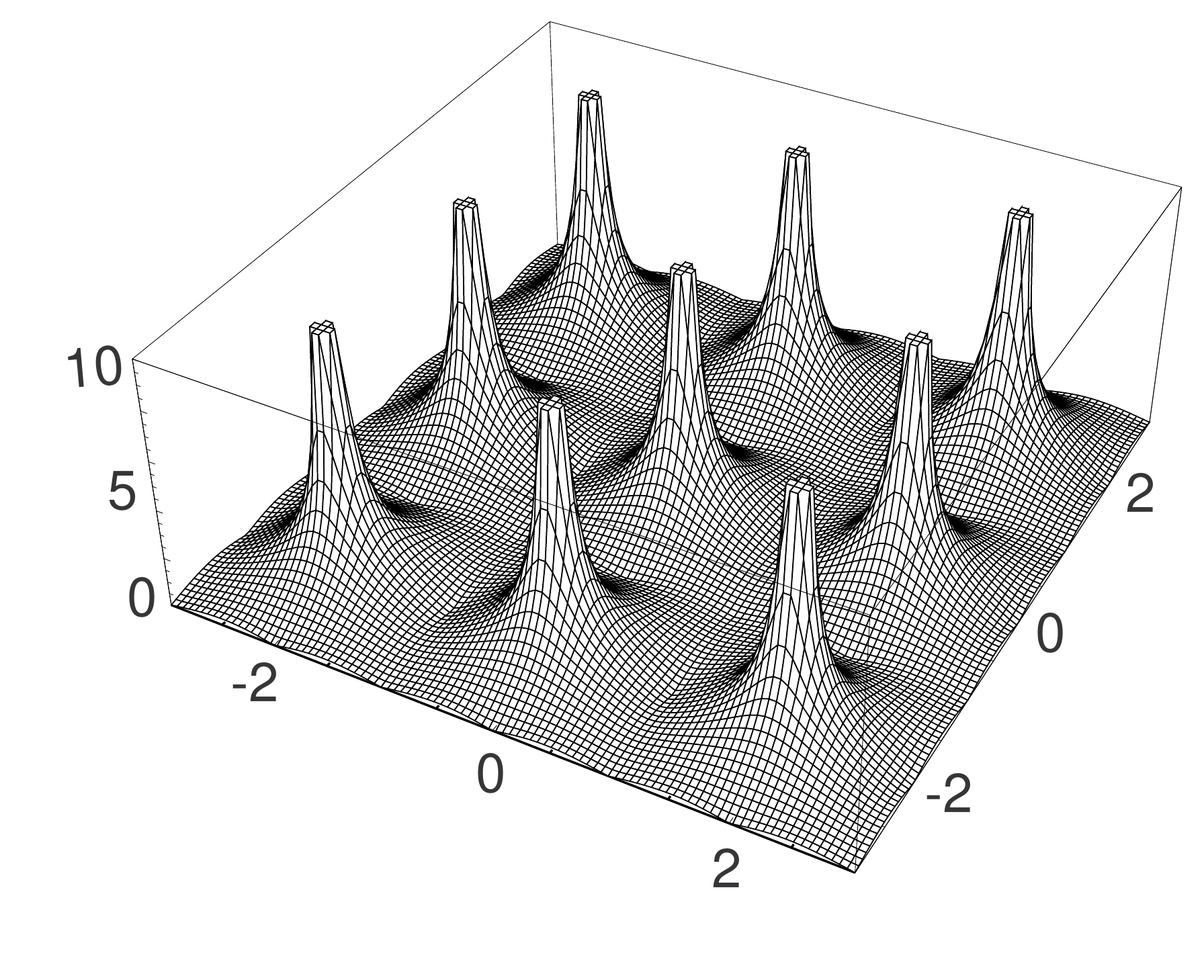}
\caption{The lattice of ordinary linear singular vortices. The distribution
of the absolute value of the superfluid velocity relative a normal component 
$\left\vert \mathbf{v}_{s}-\mathbf{v}_{n}\right\vert $ in the plane
orthogonal to the rotation axis.}
\label{fig:fig1}
\end{figure}

If we assume that $\Psi \equiv \varphi +i\psi $ is an analytic function of
the variable $Z=x+iy$, then the real and imaginary parts of this function
are the velocity potential and the stream function for some possible
two-dimensional irrotational fluid motion. By virtue of the Cauchy-Riemann
relations, any of these functions can be used to find the complex velocity.
We make use of the expression%
\begin{equation}
\Phi _{0}=\Im \ln \sigma \left( Z;g_{2},g_{3}\right) ,  \label{Fi0}
\end{equation}%
thus obtaining the complex velocity in the form%
\begin{equation}
i\zeta \left( x-iy\right) = -i\partial _{x}\Phi _{0}+\partial _{y}\Phi _{0},
\label{dz0}
\end{equation}%
The absolute value of $\upsilon _{r}\left( x,v\right) $ is shown in Figure %
\ref{fig:fig1}. %%%%%%%%%%%%%%%%%%
%%%%%%%%%%%%%%%%%%

This function has point singularity at each lattice site, where the entire
vorticity is concentrated in the central non-superfluid filament of radius $%
\xi \ll a$.

\section{A lattice of core-less vortices}

\label{sec:col}

The lattice of linear singular vortices, discussed in the previous section,
has been obtained in the case of $\beta =\pi /2$\ and $S=0$, when the
superfluid motion is described by the phase $\Phi (\mathbf{\varrho })$\ only.

We now remove the singular cores of the above vortex lattice by making use
the remaining freedom in the order parameter - the field of the triad $%
\mathbf{u}_{1}\mathbf{,u}_{2}\mathbf{,l}$\ with some trial functions $\beta (%
\mathbf{\varrho })$\ and $S(\mathbf{\varrho })$. For this we notice that the
matrix (\ref{Aij}) can be recast as $\mathsf{A}_{ij}=u_{i}u_{j}$, where $%
\mathbf{u}$ is the vector of the form%
\begin{gather}
\mathbf{u}=\frac{1}{2}e^{i\left( \frac{1}{2}\Phi +S\right) }\left( 1+\cos
\beta \right) \mathbf{e}_{+}  \notag \\
+e^{i\frac{1}{2}\Phi }\sin \beta ~\mathbf{e}_{0}-\frac{1}{2}e^{i\left( \frac{%
1}{2}\Phi -S\right) }\left( 1-\cos \beta \right) \mathbf{e}_{-}  \label{u}
\end{gather}%
with%
\begin{equation}
\mathbf{e}_{\pm }=\frac{1}{\sqrt{2}}\left( 1,\pm i,0\right) ,~\ \mathbf{e}%
_{0}=\left( 0,0,1\right) .  \label{e}
\end{equation}%
From this representation it is seen that where $\beta =\pi \ $or $0$ the
singularity of the vortex type in the phase function $\frac{1}{2}\Phi +S$ or 
$\frac{1}{2}\Phi -S$ does not lead to divergent $\mathbf{V}$~[see Eq. (\ref%
{V})] provided $\frac{1}{2}\Phi \pm S$ is analytic. Given this, we choose
two sets of lattice sites, where $\beta =0$ and $\pi $, that is, $\mathbf{l}$
is parallel and anti-parallel to $\boldsymbol{\omega }$, respectively: $\Phi
_{0}\left( x,y\right) $ and $\Phi _{0}\left( x+1,y+1\right) $ and suppose%
\begin{equation}
\Phi \left( x,y\right) =\frac{1}{2}\left[ \Phi _{0}\left( x+1,y+1\right)
+\Phi _{0}\left( x,y\right) \right]   \label{FFi}
\end{equation}%
and%
\begin{equation}
S\left( x,y\right) =\frac{1}{4}\left[ \Phi _{0}\left( x+1,y+1\right) -\Phi
_{0}\left( x,y\right) \right]   \label{S}
\end{equation}%
We thus obtained the superfluid velocity (\ref{V}) in the form of complex
function%
\begin{gather}
\mathcal{V}=\frac{1}{2}\left( 1+\cos \beta \right) i\zeta \left(
x-iy+1-i\right)   \notag \\
+\frac{1}{2}\left( 1-\cos \beta \right) i\zeta \left( x-iy\right) -i\omega
\left( x+iy\right) .  \label{Vs}
\end{gather}%
The Weierstrass zeta function has the following properties in a complex
plane \citep{Ryzhik}:%
\begin{equation}
\zeta \left( Z+2m\right) =\zeta \left( Z\right) +2m\zeta \left( 1\right) ,
\label{z1}
\end{equation}%
\begin{equation}
\zeta \left( Z+2ni\right) =\zeta \left( Z\right) +2n\zeta \left( i\right) ,
\label{z2}
\end{equation}%
\begin{equation}
i\zeta \left( 1\right) -\zeta \left( i\right) =i\pi /2.  \label{zz}
\end{equation}%
If we take $\zeta \left( 1\right) =\omega $ with $\omega =\pi /4$, then from
(\ref{zz}) it follows that $\zeta \left( i\right) =-i\omega $, so that the
superfluid velocity becomes a periodic function and the free energy becomes
the sum of that in one unit cell.

As was mentioned above, to determine the equilibrium state, we have to find
the periodic function $\beta \left( x,y\right) $ that minimizes free energy (%
\ref{Ft}). Since it is difficult to solve the minimization problem, we
consider the trial function%
\begin{equation}
\beta =\frac{\pi }{2}\left[ 1-\frac{1}{2}\left( \cos \pi x+\cos \pi y\right) %
\right] ,  \label{beta}
\end{equation}%
which has the lattice symmetry and is equal to $0$\ and $\pi $\ at the pole
points of the functions $i\zeta \left( x-iy\right) $\ and $i\zeta \left(
x-iy+1-i\right) $, respectively [see Eq.(\ref{Vs})]\emph{.}

\begin{figure}
\includegraphics[width=\columnwidth]{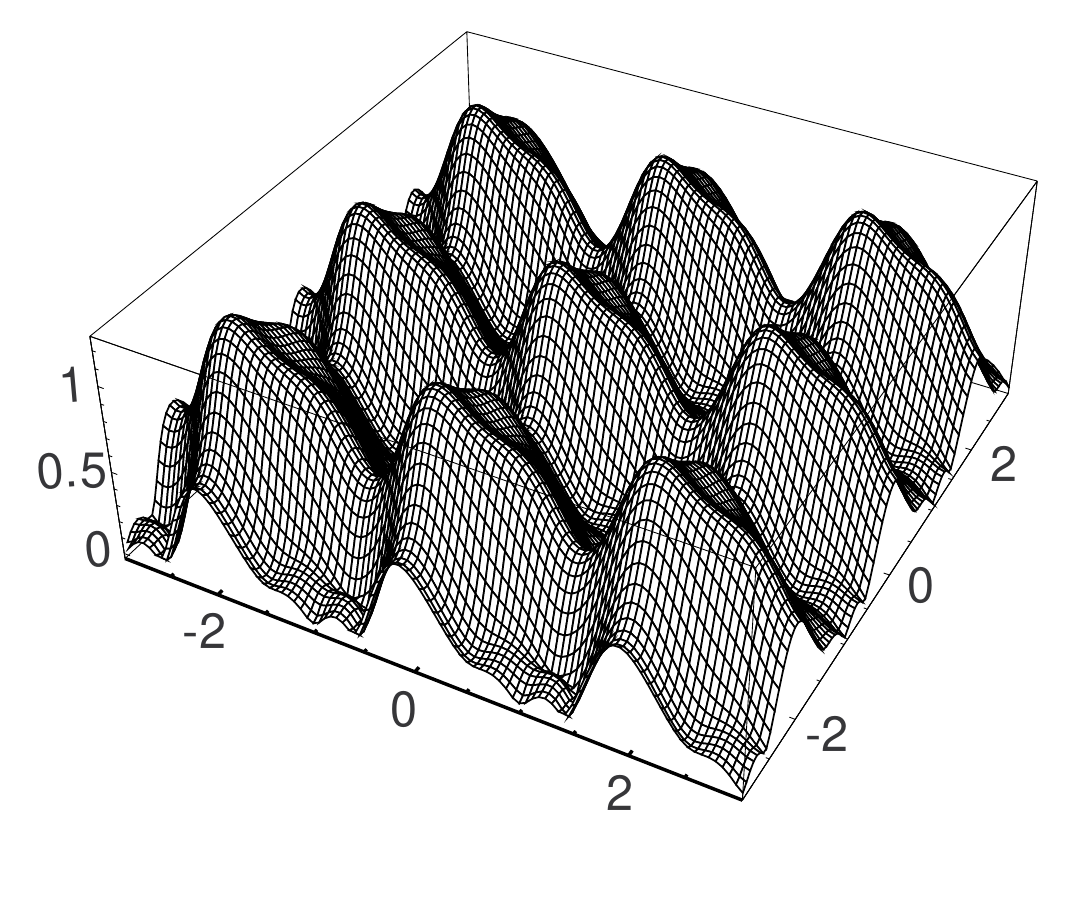}
\caption{Distribution of the absolute value of superfluid velocity in a
rotating coordinate system for a lattice of non-singular vortices with a
diffusely distributed vorticity.}
\label{fig:fig2}
\end{figure}

The resulting velocity distribution is shown in Fig. \ref{fig:fig2}, where
the absolute value of the superfluid velocity in a rotating reference frame
is depicted. %%%%%%%%%%%%%%%%%%
%%%%%%%%%%%%%%%%%%

It can be seen that the superfluid velocity is a finite periodic function
that has no singularity. To demonstrate the \textquotedblleft topographic
map\textquotedblright\ of the superfluid velocity amplitude, in Fig. \ref%
{fig:fig3}, we present a contour plot of this function, which allows us to
see the periodic lattice of continuous vortices with a vorticity distributed
over the entire unit cell.

%%%%%%%%%%%%%%%%%%
\begin{figure}
\includegraphics[width=\columnwidth]{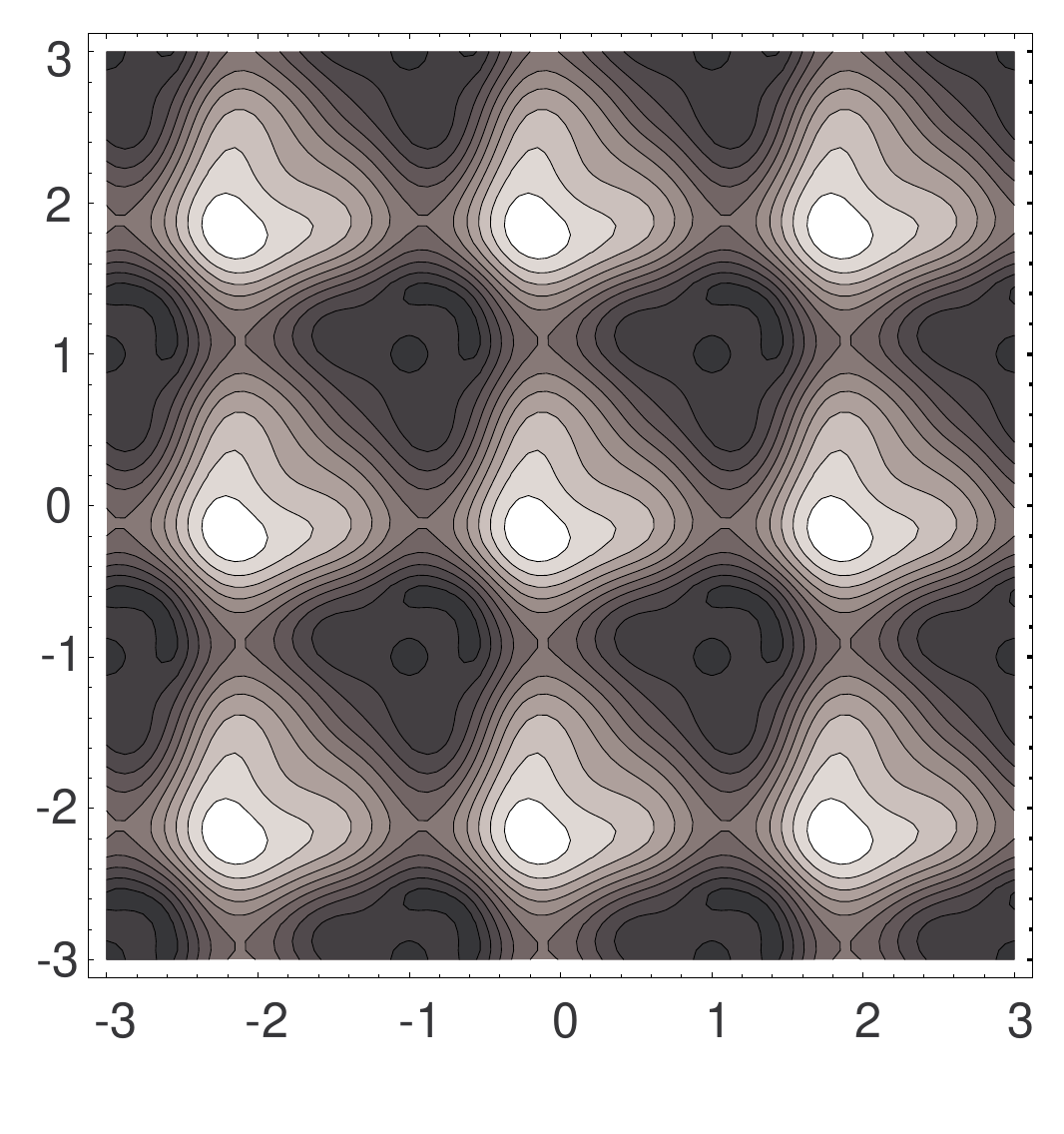}
\caption{Contour diagram of the amplitude of superfluid velocity, allowing
the reader to see the periodic lattice of continuous vortices}
\label{fig:fig3}
\end{figure}
%%%%%%%%%%%%%%%%%%

As can be seen, the modulus of the velocity $|\mathbf{v}_{s}-\mathbf{v}_{n}|$%
\ is asymmetric. This fact has a simple explanation if we recall that the $%
\mathbf{l}$\ axis indicates the direction of the quantization axis of the
total angular momentum of the pair. Apart from the kinetic term, free energy
(\ref{F})\ includes the distortion energy of the field $\mathbf{l}$\ caused
by twisting, splaying and bending, as in liquid crystals. In contrast to the
case of ordinary linear singular vortices, for a lattice of non-singular
vortices with diffusely distributed vorticity, this anisotropic contribution
is not zero. Since the total energy of the Cooper pair is conserved in the
vortex, its kinetic energy (and hence the velocity amplitude) is also
anisotropic.

Returning to the Cartesian coordinate system $x,y,z$, we use the following
simple correspondence between the complex superfluid velocity $\upsilon
(x,y) $ and the three-dimensional vector $\mathbf{v}_{s}$ in a Cartesian
system 
\begin{equation}
\mathbf{v}_{s}(x,y)=\frac{\hbar }{2Ma}\{\Re \upsilon \left( x,y\right) ,\Im
\upsilon \left( x,y\right) ,0\}.  \label{vvs}
\end{equation}%
In the laboratory frame of reference, we have%
\begin{gather}
\upsilon =\frac{1}{2}\left( 1+\cos \beta \right) i\zeta \left(
x-iy+1-i\right)  \notag \\
+\frac{1}{2}\left( 1-\cos \beta \right) i\zeta \left( x-iy\right) .
\label{vs}
\end{gather}

The vorticity of the superfluid flow is diffusely distributed over the
entire elementary cell of the vortex lattice%
\begin{equation*}
\boldsymbol{\nabla }\mathbf{\times v}_{s}=2\frac{\hbar }{2Ma}\sin \beta
\left( \boldsymbol{\nabla }S\times \boldsymbol{\nabla }\beta \right) .
\end{equation*}%
Since the superfluid velocity is concentrated in the plane of rotation, the
vorticity vector is directed along the angular velocity $\mathbf{\Omega }$\
and is continuously distributed over the lattice cells, as shown in Fig. \ref%
{fig:fig4}.

%%%%%%%%%%%%%%%%%%
\begin{figure}
\includegraphics[width=\columnwidth]{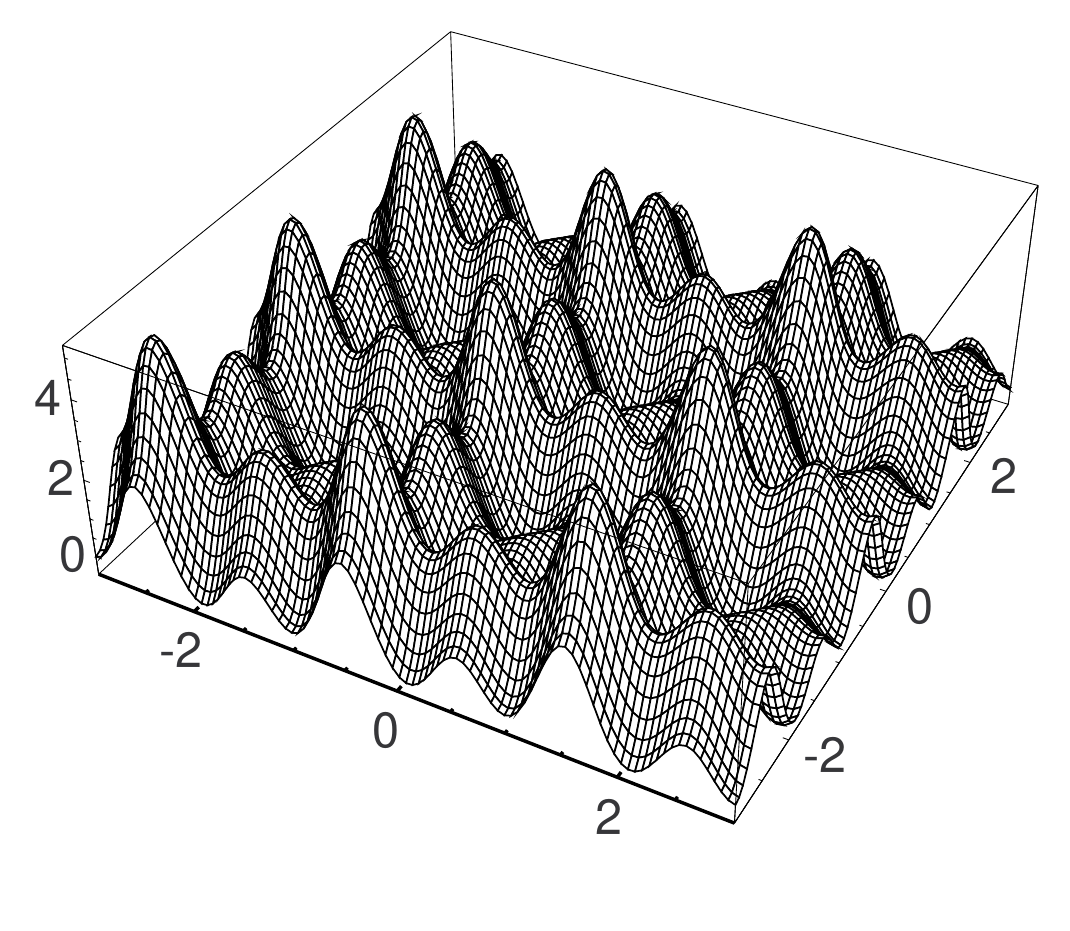}
\caption{The diffusely distributed vorticity in the rotating $^{3}$P$_{2}$
superfluid with a lattice of non-singular vortices.}
\label{fig:fig4}
\end{figure}
%%%%%%%%%%%%%%%%%%

\begin{figure}
\includegraphics[width=\columnwidth]{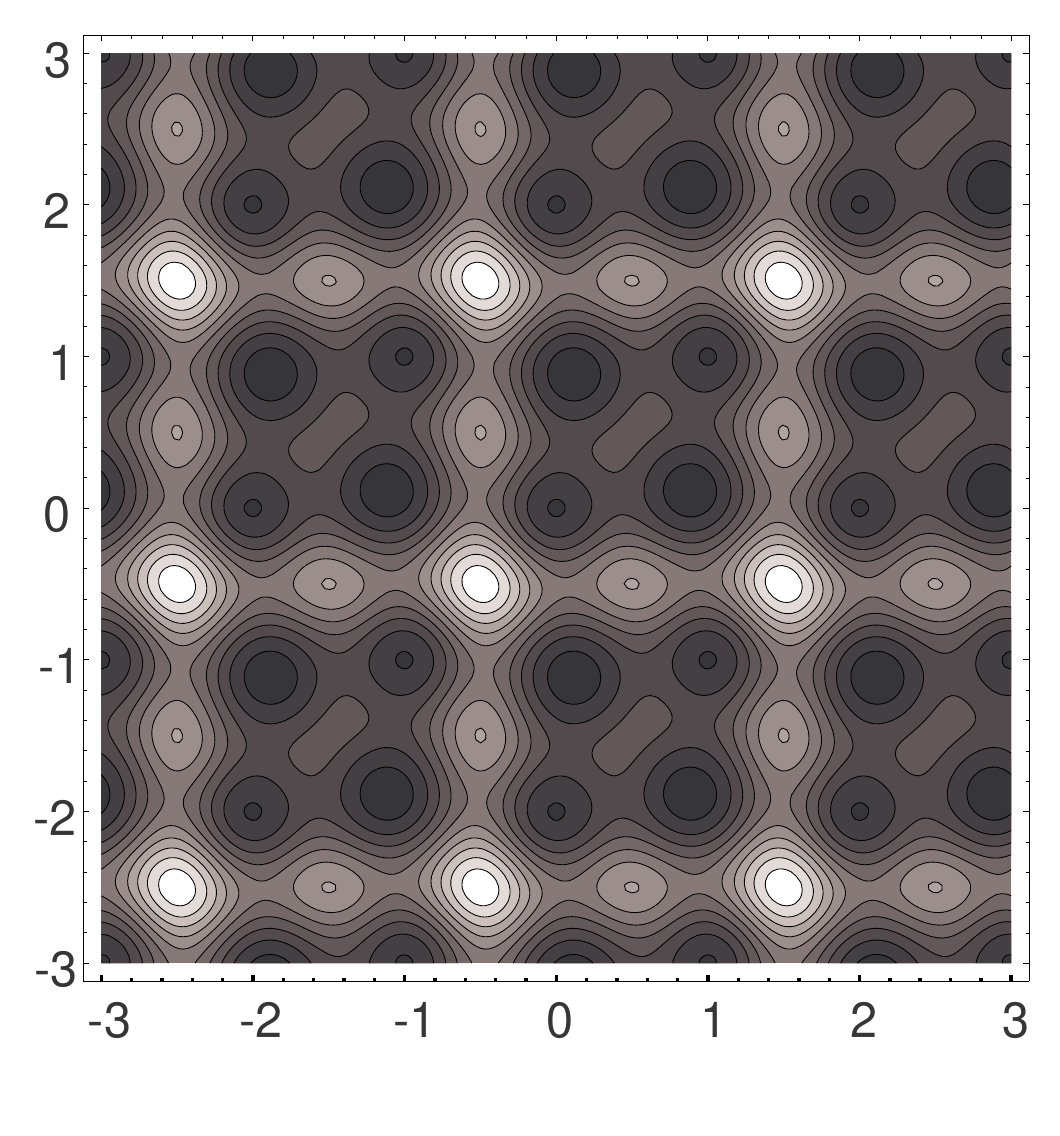}
\caption{Contour diagram of the diffusely distributed vorticity in the
rotating $^{3}$P$_{2}$ superfluid with a lattice of non-singular vortices.}
\label{fig:fig5}
\end{figure}

In Fig. \ref{fig:fig5}, we present a contour plot of this function, which
allows us to see the "topographic map" of the vorticity continuously
distributed over each unit cell.

%%%%%%%%%%%%%%%%%%
%%%%%%%%%%%%%%%%%%

Numerical integration over the cell area gives%
\begin{equation}
\int_{-a}^{a}dx\int_{-a}^{a}dy~\left( \boldsymbol{\nabla }\mathbf{\times v}%
_{s}\right) _{z}=2\pi ,  \label{kapa}
\end{equation}%
which means that each of the cells contains a unit of the quantized
vorticity $\varkappa =1$, distributed over its area. A one circulation
quantum equal to $\pi \hbar /M$, therefore the vortex density must be equal
to $2M\Omega /\pi \hbar $ so that the average vorticity of the whole
lattice, $\left\langle \boldsymbol{\triangledown }\times \mathbf{v}%
_{s}\right\rangle =2\boldsymbol{\Omega }$, corresponds to a rigid rotation
with an angular velocity of $\boldsymbol{\Omega }$.

\section{Estimation of free energy}

\label{sec:est}

With the function (\ref{beta}) we have calculated the free energy per area
carrying unit vorticity: 
\begin{equation}
\mathcal{F}=\int_{cell}\tilde{F}\left( x,y\right) dx\,dy  \label{Free}
\end{equation}%
with $\tilde{F}\left( x,y\right) $ from Eq. (\ref{F}), where we substitute
the Cartesian superfluid velocity 
\begin{equation}
\mathbf{V}(x,y)=\{\Re \mathcal{V}\left( x,y\right) ,\Im \mathcal{V}\left(
x,y\right) ,0\},  \label{Vcart}
\end{equation}%
with $\mathcal{V}\left( x,y\right) $ from Eq. (\ref{Vs}) and 
\begin{equation}
\boldsymbol{\nabla }S=\{\Re \Lambda \left( x,y\right) ,\Im \Lambda \left(
x,y\right) ,0\},  \label{grS}
\end{equation}%
where 
\begin{equation}
\Lambda \left( x,y\right) =\frac{1}{4}\left[ i\zeta \left( x-iy+1-i\right)
-i\zeta \left( x-iy\right) \right] .  \label{LAM}
\end{equation}

Using these functions, one can numerically estimate the free energy per unit
cell, as indicated in Eq. (\ref{Free}), to obtain 
\begin{equation}
\mathcal{F}=47.7\frac{\hbar ^{2}\rho _{s}}{\left( 2M\right) ^{2}}.
\label{Fnum}
\end{equation}

Which set of vortices is the most stable for a rotating superfluid neutron
system with spin-triplet pairing? To answer this question let us compare
this result with the free energy of a lattice of ordinary linear singular
vortices, which is believed to be the most stable in the inner core of
superfluid neutron stars. As we consider the problem in the London limit,
assuming spatial variations to be slow, one has to compare only derivative
terms in the free energy. For the set of ordinary linear vortices the
derivative part of the free energy consists of only the kinetic term since
the distortion energy of the $\mathbf{l}$\ field vanishes. In this case, the
problem is reduced to that for the lattice of linear singular vortices in
the superfluid $^{4}$He, which is well studied. We quote the result of %
\citet{Tkachenko:1965}\ for the energy per elementary cell of the vortex
lattice carrying a unit vorticity:%
\begin{equation}
\mathcal{F}_{0}=\frac{\hbar ^{2}\rho _{s}}{\left( 2M\right) ^{2}}\left\{ 
\begin{array}{c}
-4.117+\pi \ln \left( d/\xi \right) \text{ \ for \ \ the \ square \ lattice,}
\\ 
-4.150+\pi \ln \left( d/\xi \right) ~\ \text{for the triangular lattice.}%
\end{array}%
\right.   \label{FTK}
\end{equation}

The number $N_{v}$ of vortices with the unit circulation created inside
rotating neutron stars can be estimated to be%
\begin{equation}
N_{v}\simeq 1.9\times 10^{19}\left( \frac{1\mathrm{ms}}{P}\right) \left( 
\frac{M^{\ast }}{900\,\mathrm{MeV}}\right) \left( \frac{R}{10\,\mathrm{km}}%
\right) ,  \label{Nn}
\end{equation}%
where $P$ is the period of the neutron star, $M^{\ast }$ is the effective
neutron mass, and $R$ is the radius of the $^{3}$P$_{2}$ superfluid. Then,
we can estimate the distance between vortices $d$ from $N_{v}\pi d^{2}=\pi
R^{2}$ to get 
\begin{equation}
d=2.\,3\times 10^{-4}\left( \frac{900\,\mathrm{MeV}}{M^{\ast }}\right)
^{1/2}\left( \frac{P}{1\mathrm{ms}}\right) ^{1/2}\left( \frac{R}{10\,\mathrm{%
km}}\right) ^{1/2}\mathrm{cm}.  \label{d}
\end{equation}%
On the other hand, the coherence length of $^{3}$P$_{2}$ superfluid is about 
\begin{equation}
\xi \simeq \frac{\hbar v_{F}}{1.75\pi k_{B}T_{c}}\simeq 1.1\times
10^{-11}\left( \frac{M}{M^{\ast }}\right) \rho _{14}^{1/3}\left( \frac{10^{9}%
\mathrm{K}}{T_{c}}\right) ~\mathrm{cm},  \label{ksi}
\end{equation}%
In these equations the coherence length $\xi $ for the neutrons is related
to the corresponding Fermi velocity, $v_{F}$, and the superfluid-transition
temperature, $T_{c}$. The neutron matter density is written in units $\rho
_{14}=\rho _{n}/10^{14}\,\mathrm{g/cm}^{3}$.

Since the distance $d$ between the vortices in the core of a neutron star is
very large compared to the radius $\xi $ of the central filament of the
vortex the logarithm in Eq. (\ref{Men}) is large, $\pi \ln \left( d/\xi
\right) \sim 60$. This allows for the conclusion that the existence of an
ordered lattice slightly changes the energy of linear singular vortices, and
the single vortex approximation also gives a reasonable result. The energy
per unit length of an isolated singular vortex in the spin-triplet
superfluid of neutrons, as derived in \citep{Mendel:1991} is 
\begin{equation}
\frac{E}{L}=\pi \rho _{s}\frac{\hbar ^{2}}{4M^{2}}\ln \frac{d}{\xi }.
\label{Men}
\end{equation}

\section{Discussion}

\label{sec:dis}

According to the results of the preceding section the lattice of the
non-singular vortices with a vorticity diffusely distributed over the entire
unit cell seems to be more energetically favourable in inner superfluid core
of rotating neutron stars. This contradicts the established opinion that the
motion of neutron $^{3}$P$_{2}$ superfluid in neutron stars is irrotational.

In the evaluation of the free energy we have used the trial functions and
hence the value obtained is the upper bound. It is reasonable to feel that
the relationship between the energies of the two structures will remain the
same in the exact calculation. Of course, there appears a question to
stability of the lattices we have considered. The analogous investigations
on helium \citep{Abrikosov:1965} have shown that all is not well in
connection with stability of lattices against long waves. The stability of
the structure and the improvement of the trial functions can be further
studied by the perturbation method.

If the lattice of non-singular vortices is indeed more favourable, this may
affect the hydrodynamics of superfluid neutron stars. As is well known, drag
and pinning effects play an important role in the current theory of pulsar
glitches \citep{als84}. For recent review see e.g. \citep{alpar:2017}.
Usually the drag on neutron vortices in the core is due to electron
scattering which depends on the magnetization of the superfluid vortices %
\citep{Muzikar:1980}. It is easily to verify that the order parameter,
defined with Eqs. (\ref{Psi}) and (\ref{Aij}), satisfies the unitary
condition. This implies that the superfluid state under consideration
retains time reversal symmetry and does not have spin polarization. In other
words the core-less $^{3}$P$_{2}$\ vortices have no magnetization and the
electron scattering off the vortices is ineffective. By the same reason
pinning with flux tubes is also impossible.

We considered vortices in neutron matter without taking into account the
superfluid proton component. It can be expected that taking into account the
entrainment effects can lead to magnetization of non-singular vortices. This
problem deserves a separate consideration. 

%%%%%%%%%%%%%%%%%%555555555

\subsection*{Data availability}

The data underlying this article are available in the article and in its
on-line supplementary material. %%%%%%%%%%%%%%%%5
\bibliographystyle{mnras}

\end{document}